\documentclass[12pt]{article}
\begin{document}
\begin{center}
{\bf{
Fermat's Principle in Curved Space-time, No Emission from
Schwarzschild Black Holes as Total Internal
Reflection and Black Hole Unruh effect}}

\bigskip
{\bf{
Soma Mitra and Somenath Chakrabarty$^\dagger$

\medskip
Department of Physics, Visva-Bharati, Santiniketan-731235, 
India\\

$^\dagger$somenath.chakrabarty@visva-bharati.ac.in
}}
\end{center}
\begin{center}
Abstract
\end{center}
Using the Fermat's principle in curved space-time with stationary 
type metric, we have obtained the speed
of light as a function of spatial coordinates and hence the corresponding refractive index. 
The whole region with space dependent gravity is divided into a number of overlapping transparent refracting media with 
varying refractive index. The refractive index is found to be increasing with the strength of gravitational field. 
Hence using the laws of refraction, we have explained the
gravitational bending of
light. Further using the conventional idea of total internal 
reflection of light while going from denser to rarer medium, in the
present scenario it
is the propagation of light from the region of ultra-strong gravitational
field to relatively 
weaker gravitational field region, we have proposed an alternative approach for no emission of any kind of
electromagnetic radiation from the surface of a classical Schwarzschild Black Hole. 
We have further noticed that for an observer in a uniformly accelerated frame,
analogous to the Unruh radiation, there can be
emission of electromagnetic waves from the event horizon of a classical black hole. This may be named as "black hole
Unruh effect".

\bigskip
Around three and half century ago Fermat proposed a principle on the
propagation of light in transparent medium. It states that a 
ray of light in passing from one point to another by way of either a number of reflecting or refracting surfaces
takes a path so that the time taken to traverse is minimum. According
to this famous principle of Pierre  De Fermat, 
the nature always acts by the shortest course \cite{R1a,R1,R3}. The Laws of reflection and refraction of light are thus explained in a 
single statement. During that time the principle was thought to be valid for plane surfaces only. 
However, the principle was found to be equally true for curved surfaces and also applicable to curved paths of light. 
The analytical form of Fermat's  Principle is expressed as
\begin{equation}
\int_{1}^{2}\mu ds = {\rm{minimum}},
\end{equation}
where $ds$ is a small element of the curve traversed by light in a medium of local refractive index $\mu(r)$. 
Later the principle was replaced by a more precise form- which is the 
law of extreme path. It is expressed as the extrema of the optical
path length. The optical path is the sum of the product of local 
refractive index and the corresponding actual path element traversed by light. Then in the
case of a number of overlapping continuous media, where the refractive index also changes continuously, the above 
analytical form may be replaced by
\begin{equation}
\delta\int_{1}^{2}\mu ds = 0
\end{equation}

During the years \rm{1912-14}, while
Einstein was collaborating with Grossmann \cite{R2}
introduced a novel idea that the path of a 
freely moving particle in a gravitational field would be a geodesic in the four dimensional curved space-time of 
Riemannian geometry. This prescription was almost identical with the famous Fermat's Principle.

To consider Fermat's Principle in curved geometry of general theory
of relativity, we
follow the formalism of Landau and Lifshitz \cite{R3} (see also \cite{R2a}). The 
mathematical form of Fermat's Principle in a static gravitational
field is then given by 
\begin{equation}
\delta\int\frac{dl}{c g_{\rm{0}\rm{0}}^{1/2}(r)}=0
\end{equation}
where $c$ is the speed of light in free space and the distance $dl$
in 3-space between two points is given by
\begin{equation}
dl^{2} = \left (-g_{\alpha\beta} +\frac{g_{\rm{0}\alpha} g_{\rm{0}\beta}}{g_{\rm0\rm0}}\right ) dx^{\alpha}dx^{\beta}
\end{equation}
which from Landau and Lifsthitz may also be expressed in the form 
\begin{equation}
dl^2= \gamma_{\alpha\beta}\rm{ds}^{\alpha}\rm{ds}^{\beta}
\end{equation}
where
\begin{equation}
\gamma_{\alpha\beta} = -g_{\alpha\beta} + \frac{g_{\rm{0}\alpha}g_{\rm{0}\beta}}{g_{\rm{0}\rm{0}}}
\end{equation}
It is quite obvious from eqn.(3) that in presence of a (static) gravitational field, light can not traverse a path with 
shortest time.
To obtain an explicit mathematical expression for Fermat's Principle, we consider an isotropic  and spherically 
symmetric type metric. The general form of which is given by
\cite{R4,R5,R6}
\begin{equation}
ds^2 = g_{00}(r)c^{2}dt^2 - g_{rr}(r)dr^2
\end{equation}
where $r$ is the spatial coordinate. We can also define the spatial metric in the form
\begin{equation}
dl^2 = g_{rr}(r)dr^2 
\end{equation}
Hence
\begin{equation}
\gamma_{ij} = g_{rr}(r)\delta{ij}
\end{equation}
Now the world line corresponding to the propagation of light (electromagnetic waves or photon in quantum picture) 
is called null geodesic and is defined by the equation 
\begin{eqnarray}
\rm{ds} = \rm{0} \\
~~\rm{which~~ gives}~~ 
g_{00}(r)c^{2}dt^2 - g_{rr}(r)dr^2 = \rm{0} \\
~~\rm{Hence}~~ 
c_{g}(r) = \frac{dr}{dt} = \frac{cg_{00}^{1/2}(r)}{g_{rr}^{1/2}(r)}
\end{eqnarray}
where $c_g(r)$ is the speed of light in presence of a stationary
gravitational field.
If both $g_{\rm{00}} = 1$ and $g_{rr}(r)=1$ we get back 
the special theory of relativity result with constant value of the speed of light in vacuum, which is $c$. The speed 
of light $c_g(r)$ is therefore depends on the spatial coordinate $r$
through the space dependent metric elements. The propagation of light in presence of 
gravitational field, which is different at different spatial point is
equivalent to the motion in overlapping refracting media.
The refractive index, which depends on the strength of
gravitational field also changes from point to point. 
Therefore in curved space, light can not travel along a straight
line, it should be a geodesic.
Now using eqns.(8)-(12) the Fermat's Principle can be re-written in the form
\begin{eqnarray}
\delta\int_{1}^{2}\frac{g_{rr}^{1/2}(r)}{cg_{00}^{1/2}(r)}dr = 0
=\delta\int_{1}^{2}\frac{dr}{c_g(r)} \\
~~\rm{hence}~~ \rm{we ~~have}~~ 
\frac{c}{c_g(r)} = \frac{g_{rr}^{1/2}(r)}{g_{00}^{1/2}(r)} = n(r),
\end{eqnarray}
the refractive index of the medium, generated by the space dependent gravitational field. As we shall see in our 
subsequent discussion that the refractive index is an increasing
function of the strength of gravitational field. The 
radius of curvature for the curved path traversed by light is related to the space dependent refractive index by the 
relation \cite{R7}
\begin{equation}
\frac{1}{R} = \hat{N}.\nabla\ln\,n
\end{equation}
where $\hat{N}$ is an unit vector along the principal normal. It is
very easy to verify that this relation will give the bending of rays in the 
direction of increasing refractive index, or in the direction of increasing gravitational field. 

For the sake of illustration let us consider the Schwarzschild
metric. For simplicity, we assume that the space is
spherically symmetric. The resulting line element is then given by
\begin{eqnarray}
ds^2 &=& \left (1 - \frac{2GM}{c^{2}r}\right )c^2dt^2 - \left (1 - \frac{2GM}{c^{2}r}\right )^{-1}dr^2 \\
&&~~\rm{hence} ~~
g_{00}(r) = 1- \frac{2GM}{c^{2}r} \\
&&~~\rm{and}~~
 g_{rr}(r) = \left (1- \frac{2GM}{c^{2}r}\right )^{-1} \\
&&\rm{Hence ~the ~speed~ of ~light ~and ~the ~refractive ~index~are
~given ~by}~~\nonumber \\
c_g(r)&=&c\left ( 1-\frac{2GM}{c^2r}\right ) \\
{\rm{and}}~~ n(r) &=& 
\frac{1}{\left (1 - \frac{2GM}{c^{2}r}\right )^{1/2}}.\frac{1}{\left (1 - \frac{2GM}{C^{2}r}\right )^{1/2}} 
= \left (1-\frac{2GM}{c^{2}r}\right )^{-1}
\end{eqnarray}
It is therefore quite obvious that in presence of gravitational
field, the refractive index $n(r)> 1$ and increases with the increase in the magnitude of gravitational field strength, given by $GM/r^2$.
Which therefore also explains the phenomena of gravitational bending of light rays from the simple concept of geometrical optics. The rays will 
bend towards the gravitating object. Which further means the bending
of light towards the normal in denser medium- the conventional
picture of refraction of light. The gravitational lensing in some
sense is then equivalent to the apparent depth / distance in the space of
varying gravitational refractive index.

Let us now consider a Schwarzschild black hole. Near the event horizon 
\begin{equation}
r \approx R_s=
\frac{2GM}{c^2},
\end{equation}
where $R_s$ is the Schwarzschild radius \cite{R6}. Obviously the gravitational
field is infinitely large near the event horizon (even if we assume
the blue shifted form). 
As a consequence the gravitational refractive index will also be
infinitely large near the event horizon (this is equivalent to say that the speed of light is tending to zero, or in
other ward, the light is almost stopped). Then from the 
concept of total internal reflection of geometrical optics, 
a ray of light will be reflected internally while propagating from a medium of extremely high refractive index  
to a medium whose refractive index is low enough. In the present
scenario it is the propagation from ultra-strong 
gravitational field region near the event horizon to low
gravitational field region. In other wards, there will be total 
internal reflection of the rays near the event horizon of a black hole. 
Not a single light ray can come out from the region which is very close to the event horizon. 
Hence we may think that it is an alternative approach to explain the
phenomena that light cannot escape from the surface of
a classical Schwarzschild black hole. Further, it is well known that
the critical angle $\theta_r$ for grazing emergence, when light
travels from denser to rarer medium is given by
\begin{equation}
\theta_r=\sin^{-1}\left ( \frac{1}{n(R_s)\longrightarrow
\infty}\right ) \approx 0
\end{equation}
Therefore in the present scenario there is no allowed window for the emission of radiation near the event horizon.

Next we consider a frame undergoing a uniform accelerated motion
otherwise in flat space-time geometry \cite{R5,R8,R9,R10,R11,R12,R12a,R12b,R12c}.
Let us now study the variation of gravitational refractive index
with spatial coordinates when observed from  
uniformly accelerated frame. We assume that the frame is moving radially with a constant acceleration $\alpha$ (the
space is assumed to be isotropic in nature). 
Then according to the principle of equivalence, we can assume it to
be a rest frame in presence of a constant 
gravitational field $\alpha$. In this case the metric, known as the Rindler metric is given by 
\begin{equation}
ds^2 = \left (1+ \frac{\alpha r }{c^2}\right )^{2}c^2dt^2 - dr^2
\end{equation}
Then in this Rindler space, considering the null geodesic for 
the propagation of electromagnetic waves, the speed of light may be
written as
\begin{equation}
c_g(r)=c\left ( 1+\frac{\alpha r}{c^2}\right )
\end{equation}
Hence the refractive index is given by
\begin{equation}
n(r) =\frac{c}{c_g(r)}= \frac{1}{1+\frac{\alpha r}{c^2}}
\end{equation}
Further, it is obvious from eqns.(24) and (25) that a uniformly accelerated frame in absence of gravity can not
see the meaningfull spatial variation of the speed of light and the refractive index unless we use the
principle of equivalence.  Then substituting
\begin{equation}
\alpha=-\frac{GM}{r^2},
\end{equation}
in eqns.(24) and (25), where $M$ is the mass of the gravitating object, it is very easy to
verify that in the expressions for both the speed of light and the
refractive index there is a difference in multiplicative factor by 
$2$ from the
Schwarzschild geometry. This is solely because of the use of
principle of equivalence. It is also quite obvious that $\alpha$ can
not be a constant throughout the whole space. It remains constant within a
limited region. Then $\alpha$ may be called as local gravitational field $\alpha_l$  for the rest frame.
Now with $R_s=2GM/c^2$, the Schwarzschild radius, we have from
eqns.(24) and (25)
\begin{eqnarray}
c_g(r)&=&c\left (1+\frac{\alpha_l r}{c^2}\right ) =c\left
(1-\frac{2GM}{c^2r}\right )= c\left (1-\frac{R_s}{r}\right )\\
{\rm{and}} ~n(r) &=& \frac{1}{1+\frac{\alpha_l r}{c^2}}=\frac{1}{1-\frac{GM}{rc^2}} =\frac{1}{1 -\frac{R_s}{2r}} 
\end{eqnarray}
Here we have assumed that $M$ is the mass of the black hole of Schwarzschild radius $R_s$.
Unlike the Schwarzschild metric, here $c_g(r)\longrightarrow 0$ and the
corresponding refractive index
$n(r)\longrightarrow \infty$ for $r \longrightarrow R_s/2$, i.e., at
half of the Schwarzschild radius from the centre. However, this
region is not accessible to any observer from outside the event
horizon. Whereas near the event horizon $r\approx R_s$, $c_g(r\longrightarrow R_s)=0.5c$
and $n(r\longrightarrow R_s)=2$. 
The refractive index is therefore finite and
greater than the vacuum value ($=1$). Now in the present scenario the critical angle
$\theta_r$ for grazing emergence at the event horizon is given by
\begin{equation}
\theta_r=\sin^{-1}\left ( \frac{1}{n(r=R_s)} \right )= \sin^{-1}
\left ( \frac{1}{2}\right ) =30^0
\end{equation}
Therefore in this case, for a bunch of rays emitted
from a point on the event horizon, a solid cone like window of
vertical solid angle $60^0$ through which only electromagnetic radiation can escape. To say
in other wards, for this bunch of rays from a particular point, out
of $4\pi$ solid angle, only a small window of solid angle $60^0$ is
allowed for emergence of electromagnetic radiation. However, there
are infinite number of such points on the event horizon and
consequently infinite number of such windows. A large number of them
are overlapping in nature 
through which radiation can escape. This actually means that the
event horizon of the black hole for a uniformly accelerated observer
is almost transparent for emission. This phenomenon may be compared
with Unruh effect \cite{R13,R14,R14a,R14b,R15,R16,R16a}. In Unruh effect an observer from a uniformly
accelerated frame will see radiation in inertial vacuum. However the
inertial observer moving with uniform speed 
will always see the true vacuum. The main difference between these two
phenomena is that the Unruh effect is purely quantum mechanical in
nature. An accelerated observer is always in an excited state, therefore
while it interacts with quantum vacuum in ground state will transfer some
of its energy and excites vacuum state to some higher excited states. 
This will give rise to emission of radiation from the so called inertial vacuum,
when observed from the accelerated frame.
Whereas the present scenario is purely classical in nature. The
principle of equivalence is solely responsible for such strange
phenomenon. Since one may think of a rest frame in presence of a constant
gravitational field near the event horizon, then if the
principle of equivalence is applied at the event horizon, a classical black hole will always emits
radiation. Hence it will be extremely difficult to distinguish a black hole
from a compact stellar object or indirectly speaking such exotic classical objects with old concept of no 
emission of light or any other material particles may not exist in nature.


\begin{thebibliography}{99}
\bibitem{R1a} 
Klaus Barner, "Pierre de Fermat (1601? - 1665): His life besides mathematics.". Newsletter of the European
Mathematical Society, December, pp. 12-16, (2001);
G. Mikhail Katz, David Schaps and  Steve Shnider (2013), "Almost Equal: The Method of Adequality from Diophantus to
Fermat and Beyond", Perspectives on Science 21 (3): 7750, (2013), arXiv:1210.7750.
\bibitem{R1} Feynman Lecture, Volume-II, R.P. Feynman, R.B. Leighton
and M. Sands, Addision Wesley Publishing INC., (1963).
\bibitem{R3} L.D. Landau and E.M. Lifshitz, The Classical Theory of Fields, 
Butterworth-Heimenann, Oxford, (1975).
\bibitem{R2} A. Einstein and M. Grossmann, Entwurf einer verallgemeinerten Relativittstheorie und einer Theorie der
Gravitation [Outline of a Generalized Theory of Relativity and of a Theory of Gravitation]. Zeitschrift fur
Mathematik und Physik {\bf{62}}, 225, (1913);
A. Einstein and M.  Grossmann, Kovarianzeigenschaften der Feldgleichungen der auf die verallgemeinerte
Relativitatstheorie gegründeten Gravitationstheorie [Covariance Properties of the Field Equations of the Theory of
Gravitation Based on the Generalized Theory of Relativity].
Zeitschrift fur Mathematik und Physik {\bf{63}}, 215, (1914).
\bibitem{R2a} G. Munoz and P. Jones, arXiv:1003.3022v1 [gr-qc].
\bibitem{R4} James B. Hartle, Gravity- An Introduction to Einstein's General Relativity, Addison Wesley, New York,
(2003).
\bibitem{R5} N.D. Birrell and P.C.W. Davies,
Quantum Field Theory in Curved Space, Cambridge University Press,
Cambridge, (1982);
C.W. Misner, Kip S. Thorne and
J.A. Wheeler, Gravitation, W.H. freeman and Company, New York, (1972);
W. Rindler, Essential Relativity, Springer-Verlag, New York, 1977.
\bibitem{R6} S.L. Shapiro and S.A. Teukolsky, Black Holes, White Dwarfs
and Neutron Stars, John Wiley and Sons, New York, 1983.
\bibitem{R7} L.D. Landau and E.M. Lifshitz, Electrodynamics of Continuous Media,
Butterworth-Heimenann, Oxford, (1975).
\bibitem{R8} S. De, S. Ghosh and S. Chakrabarty, Astrophys and Space
Sci., 360:8, DOI 10,1007/s10509-015-2520-3 (2015).
\bibitem{R9} S. De, S. Ghosh and S. Chakrabarty, Mod. Phys. Lett.,
{\bf{A30}}, 1550182, (2015).
\bibitem{R10} C.G. Huang and J.R. Sun, arXiv:gr-qc/0701078, (2007).
\bibitem{R11} Domingo J Louis-Martinez, Class. Quantum Grav.,
{\bf{28}}, 036004, (2011). 
\bibitem{R12} D. Percoco and V.M. Villaba, Class. Quantum Grav.,
{\bf{9}}, 307, (1992).
\bibitem{R12a} Nicola Vona, Master Thesis in Physics, University of
Naples Federico II, (2007).
\bibitem{R12b} M. Socolovsky, arXiv:gr-qc/1304.2833.
\bibitem{R12c} G.F. Castillo Torres del, and C.L. Perez
Sanchez, Revista Mexican De Fisika, {\bf 52}, 70 (2006).
\bibitem{R13} W.G. Unruh,  Phys. Rev. {\bf{D14}}, 4, (1976).
\bibitem{R14} W.G. Unruh,  Phys. Rev. {\bf{D14}}, 870, (1976).
\bibitem{R14a} L. Acedo and M.N. Tung, anXiv:1505.07255v1 [gr-qc]
\bibitem{R14b} R. Schutzhold, arXiv:1110.6064v1 [quant-ph]
\bibitem{R15} S.W. Hawking,  Nature, {\bf{248}}, 30, (1974).
\bibitem{R16} S.W. Hawking,  Comm. Math. Phys.  {\bf{43}}, 199, (1975).
\bibitem{R16a} J. Schwinger, Phys. Rev. {\bf{82}}, 664, (1951).
\end{thebibliography}
\end{document}